\documentclass[runningheads,a4paper]{llncs}

\usepackage{graphicx}
\usepackage{float}
\usepackage{url}
\usepackage{amsfonts}
\usepackage{makeidx}
\usepackage{algorithm}
\usepackage{algpseudocode}
\usepackage{subfig}
\usepackage{multirow}
\usepackage{eurosym}
\usepackage{booktabs}
\usepackage{listings} 

\usepackage{url}
\urldef{\mailsa}\path|{Claus.Dabringer, Johann.Eder}@aau.at|
\newcommand{\keywords}[1]{\par\addvspace\baselineskip
\noindent\keywordname\enspace\ignorespaces#1}

\begin{document}

\newtheorem{mydef}{Theorem}

\mainmatter  

\title{Adaptive Distributed Top-$k$ Query Processing}


%
%
\author{Claus Dabringer \and Johann Eder
\thanks{The work reported here was supported by the Austrian
Ministry of Science and Research within the program GENAU
(project GATIB II) and within the project BBMRI.AT.}%
}
\authorrunning{Adaptive Distributed Top-$k$ Query Processing}

\institute{Alpe Adria University Klagenfurt, Department of Informatics Systems\\
\mailsa\\}

%
%

\toctitle{Lecture Notes in Computer Science}
\tocauthor{Authors' Instructions}
\maketitle

\begin{abstract}
 ADiT is an adaptive approach for processing distributed top-$k$ queries over peer-to-peer networks optimizing both system load and query response time. This approach considers the size of the peer to peer network, the amount $k$ of searched objects, the network capabilities of a connected peer, i.e. the transmission rate, the amount of objects stored on each  peer, and the speed of a  peer in processing a local top-$k$ query. In extensive experiments with a  variety of scenarios we could show that ADiT outperforms state of the art distributed query processing techniques.
\end{abstract}

\section{Introduction}
\label{chap:ADiT-sec:Intro}
Top-$k$ queries retrieve the $k$ tuples of a query result which score best for a given objective function. Top-$k$ queries help to overcome the problem of too large query results on one hand and too low recall, if the query is more constrained, and are therefore a promising technology for improving and accelerating search in for various data collections, e.g. for the search for suitable samples in biobanks \cite{Eder:2009:ISF:1616930.1616937}, our main application area. Top-$k$ queries are also popular for providing users  ranked search results they are used from web search engines. Top-$k$ queries, in particular, the optimization of top-$k$ query processing for central databases received a lot of attention \cite{BPA2,Bruno:2002:TKS:568518.568519,Dabringer:2011:ETR:1982185.1982414,Dabringer:2011:FTQ:2033546.2033563,962155,1325952,1391730,1272749,MS_TopK,OPT}.
Optimizing top-$k$ queries in distributed environment, in particular in highly distributed networks of federated or peer to peer databases still has significant research needs.

Current distributed top-$k$ query processing approaches either focus on reducing the amount of transmitted queries \cite{Akbarinia:2006:RNT:1136637.1136639,Hagihara:2009:MPM:1590953.1590977} or on keeping the amount of transported objects low \cite{Balke2005,Conner2007,Fang:2010:BPA:1917832.1918852,Ryeng:2011:EDT:1997251.1997277}, or to reduce the communication costs{\cite{fang2014efficient}. However,  both the transmitted objects and messages affect the \textit{system effort} and \textit{query response time} in a peer to peer system. Therefore, we introduce an adaptive distributed top-$k$ (short ADiT) query processing approach considering both. To the best of our knowledge ADiT is the first approach using the amount of messages \textit{and} the amount of transmitted objects to measure \textit{system effort} and \textit{query response time}.

Processing a top-$k$ query in a p2p network with horizontal partitioning involves sending a top-$k$ query to each peer. The optimization problem now is to determine a proper $k_p$ for each $p$ of the peers, i.e. how many objects should be fetched from which peer. If this $k_p$ is too large, it results in unnecessary computation at the peers' site and unnecessary traffic. If it is too low, it is necessary to send additional queries to the peers.

In our approach several parameters are used to calculate a proper $k_p$:
\begin{itemize}
	\item size of the peer to peer network
	\item amount k of searched objects
	\item network capabilities of each  peer, i.e. the transmission rate
	\item amount of objects stored on each connected peer
	\item speed of a  peer, i.e. the searching performance of that peer
\end{itemize}



In the following  we show the general architecture for processing top-$k$ queries in a p2p environment and derive some heuristics based on the parameters outlined above. We describe the implementation of the ADiT approach and show in an extensive set of experiments the performance gains using this approach.

\section{ADiT in General}
\label{chap:ADiT-sec:General}
 \underline{A}daptive  \underline{Di}stributed  \underline{T}op-K query processing (short ADiT) is able to process distributed top-$k$ queries over horizontally partitioned data exactly. ADiT assumes a dynamic peer to peer network. Each peer has variable bandwidth capabilities and individual  message costs. In contrast to other approaches \cite{Ryeng:2011:EDT:1997251.1997277,Vlachou:2008:ETQ:1376616.1376692} ADiT does not rely on caching techniques. Thus the performance is not dependent on stable data or on reoccurring queries.

The aim of ADiT is to achieve a low overall system effort as well as a fast query response time. The first parameter, the overall system effort is defined as sum all amounts of time of the peers needed for (1) sending requests to other peers in the network to obtain further objects, (2) searching objects and (3) transmitting objects. The second parameter is the query response time, the time elapsed between submitting a query and the return of the result. Formula \ref{chap:ADiT-eq:syseffort} and formula \ref{chap:ADiT-eq:queryAnswerTime} define the system effort, respectively the query response time where $MsgCount_i$ is the total amount of messages sent to peer $P_i$ and $n_i$ is the amount of objects retrieved from peer $P_i$. We use the following abbreviations throughout of this paper:  N is the peer to peer network,  Q is the top-$k$ query, R is the queried relation, and $P_i$ is a peer in the peer to peer system.

\begin{eqnarray}
\label{chap:ADiT-eq:syseffort}
SE(N, Q, R) & = & \sum_{i=1}^{|P|} CCN.P_i, MsgC_i) +\\
& & DBCosts(N.P_i, Q, R, n_i) + \nonumber \\
& & TransCosts(N.P_i, R, n_i) \nonumber
\end{eqnarray}

\begin{eqnarray}
\label{chap:ADiT-eq:queryAnswerTime}
QueryAnswerTime (N, Q, R) & = & max(CommCosts(N.P_i, MsgCount_i),\\
& & DBCosts(N.P_i, Q, R, n_i), \nonumber \\
& & TransCosts(N.P_i, R, n_i)) \nonumber
\end{eqnarray}

The unit of system effort as well as of query response time is seconds. Thus it is needed to map the different costs to a time factor. Function \ref{chap:ADiT-eq:CommunicationCosts} defines how sending $MsgCount$ requests to peer $P$ is mapped to a time factor. The amount of incoming messages is multiplied with the constant costs that arise when establishing a connection to peer $P$. This gives the amount of time that is spent by sending $MsgCount$ messages to peer $P$.

\begin{eqnarray}
\label{chap:ADiT-eq:CommunicationCosts}
CommCosts (P, MsgCount) = P_{MsgCosts} * MsgCount
\end{eqnarray}

Function \ref{chap:ADiT-eq:TransmissionCosts} defines how retrieving $n$ objects from relation $R$ of peer $P$ is mapped to a time factor. The transmission costs are influenced by the size of the object in relation $R$ on peer $P$ and by the transmission rate of peer $P$.

\begin{eqnarray}
\label{chap:ADiT-eq:TransmissionCosts}
TransCosts (P, R, n) = \frac{(P_{R_{ObjectSize}} * n)} {P_{TransRate}}
\end{eqnarray}

The database costs ($DBCosts(N.P_i, Q, R, n)$) for searching the best $n$ objects in relation $R$ on peer $P_i$ strongly depend on the
top-$k$ approach used on peer $P_i$, performance of the answering peer $P_i$, and the issued query $Q$, e.g. on the number of restrictions.
ADiT assumes that each peer provides an estimate  of the time needed to return the \textit{top-$k$} objects for a query with \textit{m} restrictions on a relation with size \textit{N}. There is no assumption which procedure a peer uses to process top-$k$ queries..

ADiT works iteratively and calculates a separate fetch size $k'_p$ for each peer in each iteration.
Then ADiT broadcasts the  query $Q$ \textit{in parallel} and gathers the top-$k'_p$ from each peer $p$. Then ADiT tries to publish objects and repeats if necessary.

There are two major possibilities for tuning: Choosing an appropriate fetch size $k'_p$ for each peer in each iteration and avoiding to contact peers which cannot contribute to the result. For choosing the fetch size there are two extreme cases:
	\begin{enumerate}
		\item Setting $k'_p = 1$ for each peer leads to a minimal amount of \textit{transmitted objects} but to a higher amount of \textit{transmitted messages}.
		
		\item Setting $k'_p = k$ for each peer leads to a minimal amount of \textit{transmitted messages} but to a higher amount of \textit{transmitted objects}.
	\end{enumerate}

In the rest of this paper we will focus on how to tune this basic distributed top-$k$ query processing approach.


\section{Heuristic Fetch Size Calculations}
\label{chap:ADiT-sec:LargeTestDerivingHeuristics}
Analyzing a large number of queries varying the influencing factors \cite{phddabringer} we developed two heuristics (basic and enhanced) for choosing a good fetch  size $k'_p$ for each individual peer $p$.

\noindent \textbf{Basic Heuristics.} The basic heuristics shown in equation \ref{chap:ADiT-eq:fetchSizeHeuristicBasic} only uses the amount of relevant peers $N_{Size}$ and the amount of searched objects $k$ to derive a common fetch size $f$ for all peers. The basic heuristics does not assume any particular data distribution. Thus it tries to retrieve an equal amount of objects from each peer. In case $k$ is larger than $N_{Size}$ the basic heuristics equally distributes $k$ among the available peers. Otherwise the basic heuristics calculates the smallest multiple of $k$ which is greater or equal than $N_{Size}$ and equally distributes this amount among the available peers. The $consFactor$ is used to increase the fetch size since it is unlikely that  each peer will contribute the same number of objects. This increasing is used to fetch more objects and keep the number of iterations small. Our initial experiments showed that a $consFactor$ of 2 leads to good results, e.g. few iterations and thus few messages exchanged in the p2p network. If the data is not  distributed equally, $consFactor$ should be chosen higher.

\begin{eqnarray}
\label{chap:ADiT-eq:fetchSizeHeuristicBasic}
f & = &  min(k, consFactor * \left \lceil \frac {N_{Size}} {k} \right \rceil * \frac{k} {N_{Size}}))
\end{eqnarray}

\noindent \textbf{Enhanced Heuristics.} The enhanced heuristics calculates the fetch size $k'_p$ for each peer p \textit{separately}. It uses additional parameters to adjust the fetch size for each peer properly:
\begin{itemize}
	\item $ObjectsStored_{p}$: Amount of objects stored on peer p.
	\item $ObjectsStored_{N}$: Amount of objects stored in the peer to peer system N, i.e. $sum(ObjectsStored_{p})$
	\item $Speed_{p}$: Query processing speed of peer p, e.g. a value between 1 and 10 where 1 is the slowest and 10 the fastest speed.
	\item $maxSpeed_{N}$: Maximum query processing speed of a peer in the peer to peer system N.
	\item $TransRate_{p}$: The transmission rate describing how fast the network connection of a certain peer is. This value is given in MBit per second.
	\item $maxTransRate_{N}$: Maximum transmission rate of a peer in the peer to peer system N.
\end{itemize}

The knowledge gathered during query processing iterations comprises the following parameters:
\begin{itemize}
	\item $ObjectsRetrieved_{p}$: Amount of objects of peer p which have already been retrieved, initially 0.
	\item $ObjectsPublished_{p}$: Amount of objects of peer p which made it in the top-$k$ answers, initially 0.
	\item $ObjPub_{N}$: Amount of objects returned to the user, initially 0.
\end{itemize}

All these parameters are used to calculate different weights which influence the enhanced heuristics. Applying the basic heuristics to the large test scenarios showed that the proposed fetch size should be treated as a lower limit. Therefore, the enhanced heuristics uses the different weights to \textit{increase the fetch size} determined with the basic heuristics. To accomplish that the enhanced heuristics maps its weights to the interval of [1, 2]. This prevents from fetching fewer objects than the basic heuristics suggested. The enhanced heuristics assumes that all previous iterations can be used to reason about following iterations, e.g. it assumes that peers that contributed more objects in previous iterations will also contribute more objects in the following iterations. This assumption is reflected in weight $w_{pF}$ which is defined in equation \ref{chap:ADiT-eq:weightPubFrac}. The more objects a peer published compared to all other peers, the more objects are gathered from this peer \textit{in the next iteration}.
\begin{eqnarray}
\label{chap:ADiT-eq:weightPubFrac}
w_{pF} = (1 + \frac{ObjectsPublished_{p}} {ObjPub_{N}})
\end{eqnarray}

The enhanced heuristics tries to reduce the amount of fetched objects which are not needed. Thus it fetches more objects from peers where the ratio between fetched objects and published objects is high. Equation \ref{chap:ADiT-eq:weightUsedFrac} shows the definition of weight $w_{uF}$.
\begin{eqnarray}
\label{chap:ADiT-eq:weightUsedFrac}
w_{uF} = (1 + \frac{ObjectsPublished_{p}} {ObjectsRetrieved_{p}})
\end{eqnarray}

The enhanced heuristics assumes that peers which store more objects will contribute more to the final answer. Thus it suggests to fetch more objects from larger peers. It uses equation \ref{chap:ADiT-eq:weightDBFrac} to incorporate that fact, namely weight $w_{DBF}$.
\begin{eqnarray}
\label{chap:ADiT-eq:weightDBFrac}
w_{DBF} = (1 + \frac{ObjectsStored_{p}} {ObjectsStored_{N}})
\end{eqnarray}

Since it is cheap to ask a faster peer for more objects the enhanced heuristics defines $w_{Speed}$ and $w_{TransRate}$. Equation \ref{chap:ADiT-eq:weightSpeed} models the fact that more objects should be fetched from peers which are faster in searching their databases.
\begin{eqnarray}
\label{chap:ADiT-eq:weightSpeed}
w_{Speed} = (1 + \frac{Speed_{p}} {maxSpeed_{N}})
\end{eqnarray}

Equation \ref{chap:ADiT-eq:weightTransRate} deals with the transmission of objects. It reflects that more objects should be fetched from peers which have a higher transmission rate.
\begin{eqnarray}
\label{chap:ADiT-eq:weightTransRate}
w_{TransRate} = (1 + \frac{TransRate_{p}} {maxTransRate_{N}})
\end{eqnarray}

The weights described in equations \ref{chap:ADiT-eq:weightPubFrac}-\ref{chap:ADiT-eq:weightTransRate} are used by the enhanced heuristics to influence the basic heuristics. The weighted fetch size is determined with the heuristic function shown in equation \ref{chap:ADiT-eq:fetchSizeHeuristicWeights}.

\begin{eqnarray}
\label{chap:ADiT-eq:fetchSizeHeuristicWeights}
k'_{p} & = &  min(k - ObjPub_{N}, \left \lceil f * w_{pF} * w_{uF} * w_{DBF} * w_{Speed} * w_{TransRate} \right \rceil) \nonumber \\
& &
\end{eqnarray}

The upper bound for fetch size $k'_{p}$ is obviously the amount of missing objects, namely $k - ObjPub_{N}$. The enhanced heuristics does not fetch more objects than the amount of missing objects from any of the peers in the peer to peer system.

\subsection{ADiT Processing Iterations}
\label{chap:ADiT-sec:DetailProcessingIterations}
ADiT processes a given distributed top-$k$ query through a number of iterations. Each iteration is used to gather objects from the peers within the system to satisfy the distributed top-$k$ query. In this section we focus on the relevant steps in each iteration. The pseudo-code in listing \ref{chap:ADiT-listing:ADiT} shows how ADiT obtains the best k objects for a list of restrictions.

The variables used for storing the maximum remaining score ($maxRemScore$) and all fetched objects ($fetchedObjs$) are assumed to be globally visible to all threads during execution. They are depicted as in-out parameters in all pseudo-codes where they are used. The output produced by the \textit{ADiT}-method is a sorted list of the $k$ objects which score best among all objects in the peer to peer system with respect to the objective function.

\textbf{Identify Relevant Peers.} ADiT only distributes the top-$k$ queries to \textit{relevant} peers. A peer p is relevant iff the last delivered object of peer p (i.e. the one with the maximum remaining score on peer p) is among the best k objects of already fetched objects, otherwise peer p is irrelevant and can be pruned, since peer p cannot return a better object than its last published object.  The set of relevant peers is updated in each iteration.

\textbf{Calculating Individual Fetch Sizes.} In each iteration ADiT assigns an individual fetch size $k'_{p}$ to each relevant peer p. The fetch size is determined using the enhanced heuristics discussed in section \ref{chap:ADiT-sec:LargeTestDerivingHeuristics}.

\begin{lstlisting}[caption=Pseudo-code for ADiT, label=chap:ADiT-listing:ADiT]
program ADiT (IN string tableName, IN Number k,
              IN Set<Restriction> restr,
              IN Set<Function> Sim, IN Function Obj,
              I_O Map<Number, object> objects)
   var maxRemScore: Number;
   var fetchedObjs: Map<Number, object>;
   var ObjPublished: Number;
   var relPeers: Set<Peer>;
   var t: Thread;
begin
    loop
         maxRemScore = 0;
         GetRelevantPeers(objects, I_O relPeers);
         CalcFetchSize(k - objects.count, I_O relPeers);
        -- broadcast
         foreach Peer p in relPeers
            t = new Thread();
            t.start(LocalTopKCall(tableName, restr, Sim, Obj,
                               p, fetchedObjs, maxRemScore));
         end-for;
        -- publish
         PublishObjects(I_O fetchedObjs, I_O maxRemScore, k,
                        relPeers, ObjPublished, objects);

    until ObjPublished == k
end.
\end{lstlisting}


\textbf{Broadcasting Top-K Query.} Within each iteration ADiT gathers objects to satisfy the distributed top-$k$ query. Therefore, ADiT distributes the query throughout the system and obtains $k'_{p}$ objects from each peer \textit{in parallel}.

For each relevant peer ADiT starts a separate thread ($LocalTopKCall$) which encapsulates two major tasks: (1) execution of a local top-$k$ query and (2) updating of the maximum remaining score if it changed (\ref{chap:ADiT-listing:LocalTopKCall}).

\begin{lstlisting}[caption=Pseudo-code for sending a top-$k$ query to a certain peer, label=chap:ADiT-listing:LocalTopKCall]
program LocalTopKCall (IN string tableName,
                       IN Set<Restriction> restr,
                       IN Set<Function> Sim,
                       IN Function Obj, IN Peer p,
                       I_O Map<Number, object> fetchedObjs,
                       I_O Number maxRemScore)
begin

    p.TQQA(tableName, q = 0, p.k', restr,
            searchType = AT_MOST,
            Sim, Obj, p.Objects));

    lock(fetchedObjs, maxRemScore);
        fetchedObjs.AddAll(p.Objects);

        if p.maxScore > maxRemScore then
           maxRemScore = p.maxScore;
        end-if;
    end-lock;
end.
\end{lstlisting}

The first part shows the call of a local $TQQA$ query processor \cite{phddabringer} which is reentrant, i.e. gathering k'=5 objects in the first iteration and k'=10 objects in the second iteration finally gives the best 15 objects from peer p. After the best (or even next in each following iteration) $k'$ objects have been retrieved they are added to a global buffer. Finally the maximum remaining score is updated in case peer p has a higher maximum score than all other peers.

\textbf{Publishing Objects.} The last step in each iteration is the publishing of relevant objects (Listing \ref{chap:ADiT-listing:PublishObjects}). Since ADiT is an \textit{exact distributed top-$k$ query processing approach} it is necessary to wait for all peers to return at least one result. This is indicated with the $waitForAll$ method. 
After all peers provided their results ADiT iterates over the \textit{sorted} map and tests for each object whether its score is greater or equal than the maximum remaining score. In that case an object can be published. ADiT stops when  enough objects have been published.

\begin{lstlisting}[caption=Pseudo-code for the publishing of objects in ADiT, label=chap:ADiT-listing:PublishObjects]
program PublishObjects (I_O Map<Number, object> fetchedObjs,
                        I_O Number maxRemScore, IN Number k,
                        IN Set<Peer> relPeers,
                        I_O Map<Number, object> objects)
begin
    waitForAll(relPeers);

    foreach Element e in fetchedObjs.Elements
       	 if e.Score >= maxRemScore
       	    objects[e.Score] = e.Object;
       	
       	    if objects.count == k then
       	       break;
       	    end-if
       	 else
       	    break;
       	 end-if
    end-for
end.
\end{lstlisting}

\section{Prototype and Experiments}
\label{chap:ADiT-sec:ExperimentalResults}
ADiT has been completely implemented in PL-SQL \cite{PLSQL,Feuerstein:1999:OPP:555010} as a set of stored procedures \cite{Owens:1998:BID:272975}. To compare ADiT against a state of the art distributed top-$k$ query processing technique we also implemented the \textit{algorithm with remainder top-$k$ queries (short ARTO) \cite{Ryeng:2011:EDT:1997251.1997277}} in this database layer.

\subsection{Experimental Setup}
\label{chap:ADiT-sec:ExperimentalSetup}

We performed experiments on 2 databases: One filled with randomly generated data, and the other consisting of a single relation containing 68 categorical attributes  taken from the \textit{UCI Machine Learning Repository} \cite{UCICensusData,Frank+Asuncion:2010} which contains over 2.400.000 entries in this single relation which we distributed  among the peers in the network such that the size of the database of each peer varied between 5.000 objects and 500.000 objects.

Within this section we present various diagrams generated from the data produced by the conducted test runs. We primarily focused on the \textit{system effort} caused by a certain query and on the \textit{query response time}. To make precise statements about ADiT and the enhanced heuristics we used the basic heuristics with a $consFactor$ of 2 and four other heuristics to compare them to the enhanced heuristics:

\begin{enumerate}
	\item $k'_p = k$
	\item $k'_p = 1$
	\item $k'_p = \left \lceil \frac{k}{N} \right \rceil$
	\item $k'_p = \left \lfloor \frac{k}{N} \right \rfloor$
	\item $k'_p = min(k, 2 * \left \lceil \frac {N_{Size}} {k} \right \rceil * \frac{k} {N_{Size}}))$
\end{enumerate}

For an easier comparison of the achieved results we defined two ratios:  gain with respect to system effort is defined in equation \ref{chap:ADiT-eq:RatioSysEffort};  gain achieved for the query response time is shown in equation \ref{chap:ADiT-eq:RatioQAT}. The respective ratios for the comparison with ARTO are defined accordingly.

\begin{eqnarray}
\label{chap:ADiT-eq:RatioSysEffort}
Ratio_{SE} & = & \frac{SystemEffort_{heuristic_i}}{SystemEffort_{heuristic_{enhanced}}}
\end{eqnarray}

\begin{eqnarray}
\label{chap:ADiT-eq:RatioQAT}
Ratio_{QAT} & = & \frac{QueryAnswerTime_{heuristic_i}}{QueryAnswerTime_{heuristic_{enhanced}}}
\end{eqnarray}

\subsection{Discussion of Results}

In 
figure \ref{chap:ADiT-fig:ratio_qat_19peers_04cons_cens} we can see the $Ratio_{QAT}$ for a query with 4 restrictions.  Comparing with figure \ref{chap:ADiT-fig:ratio_qat_49peers_04cons_cens} we can see that all curves get  higher in a peer to peer network with 49 peers. Additionally, these first figures already show that the heuristics $k'_p = 1$ is not a good choice since it involves high interaction between the query initiator and the other peers. We can also observe that for the query response time the gain over ARTO is rapidly increasing when the amount of searched objects increases. The ratio is growing fast because ARTO needs more sequential message processing when the search amount increases (when the first parallel call was not sufficient).

\begin{figure*}[!ht]
\centering

	\parbox{10cm}{
		\includegraphics[width=10cm]{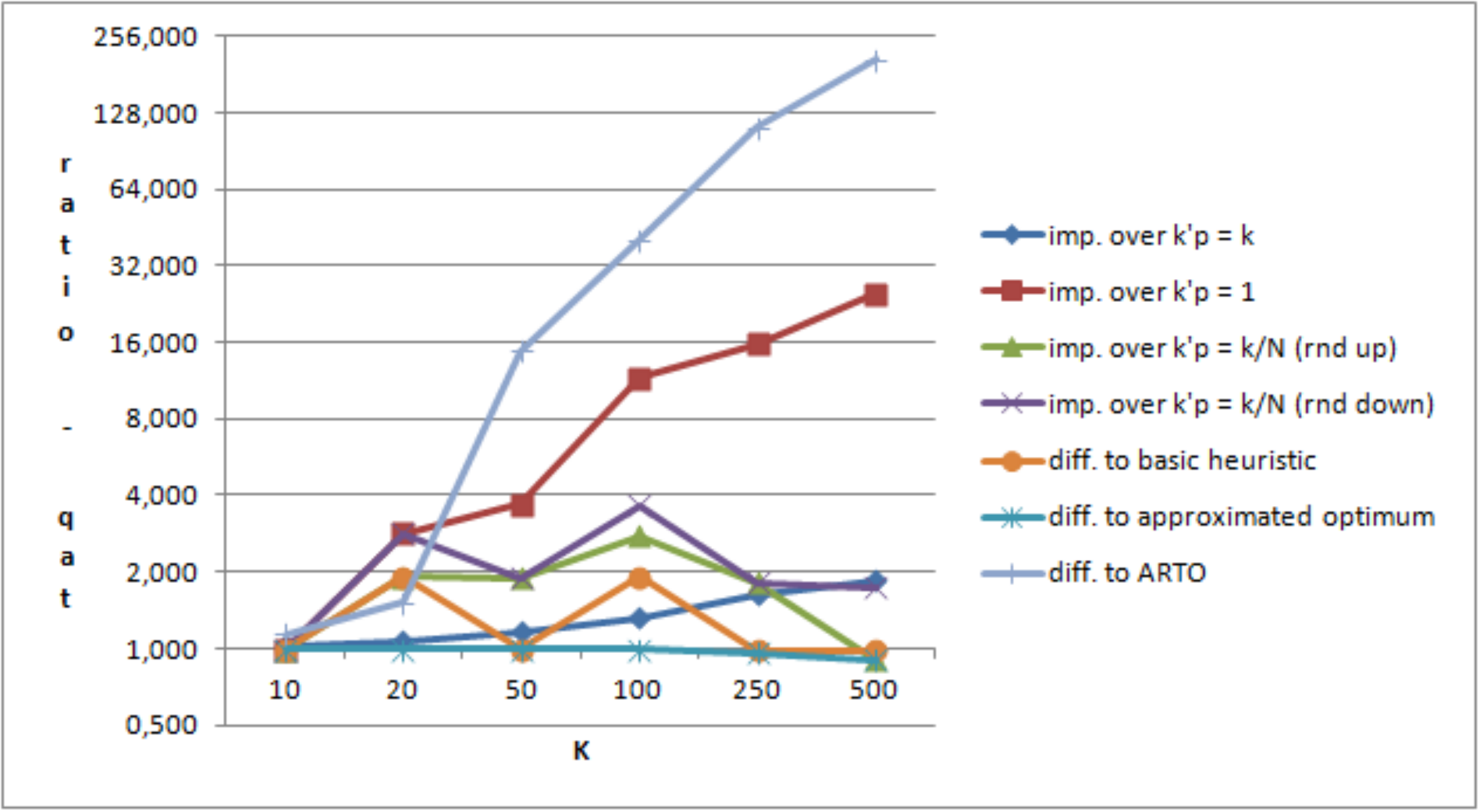}
	\caption{Ratio for query response time between \textit{enhanced heuristics}, approximated optimum, ARTO and five different approaches to determine the fetch size $k'_p$ in a peer to peer system with 19 peers and varying search amount K and 4 restrictions on census data.}%
	\label{chap:ADiT-fig:ratio_qat_19peers_04cons_cens}}%
\end{figure*}



In figure \ref{chap:ADiT-fig:ratio_sysEff_49peers_04cons_cens} and figure \ref{chap:ADiT-fig:ratio_qat_49peers_04cons_cens} we can see  $Ratio_{SE}$ and  $Ratio_{QAT}$ for a query with 4 restrictions in a peer to peer network with 49 peers storing census data. We can observe that the ratio $Ratio_{SE}$ and $Ratio_{QAT}$ are almost identical with respect to their curves. They only differ in the magnitude which is a little higher for the $Ratio_{QAT}$. This means that the usage of ADiT brings slightly more benefits to a single user than to the whole peer to peer system. This result can be observed over all of the tests. The reason for this behaviour is that ADiT tries to fetch fewer objects from less important peers. Thus these peers do not influence the search process that much than in a setting where all peers are contributing the same amount of objects. Another reason is that the search time is dominated by the slowest peer. Avoiding high interaction and fetching few objects from such peers can clearly boost query processing. Another observation is that ARTO has a lower system effort for a small search amount. This can be seen in figure \ref{chap:ADiT-fig:ratio_sysEff_49peers_04cons_cens}. The reason for this is that ARTO can answer queries with fewer  messages and fewer transmitted objects. This is because ARTO sequentially asks the peer with the highest remaining score for further objects which results in fewer work for the remaining peers. However in figure \ref{chap:ADiT-fig:ratio_qat_49peers_04cons_cens} we can observe that the query response time is better for ADiT in the same scenario.

\begin{figure*}
\label{chap:ADiT-fig:ratios_49peers_04cons_cens}
\centering
	\begin{minipage}{10cm}
		\includegraphics[width=10cm]{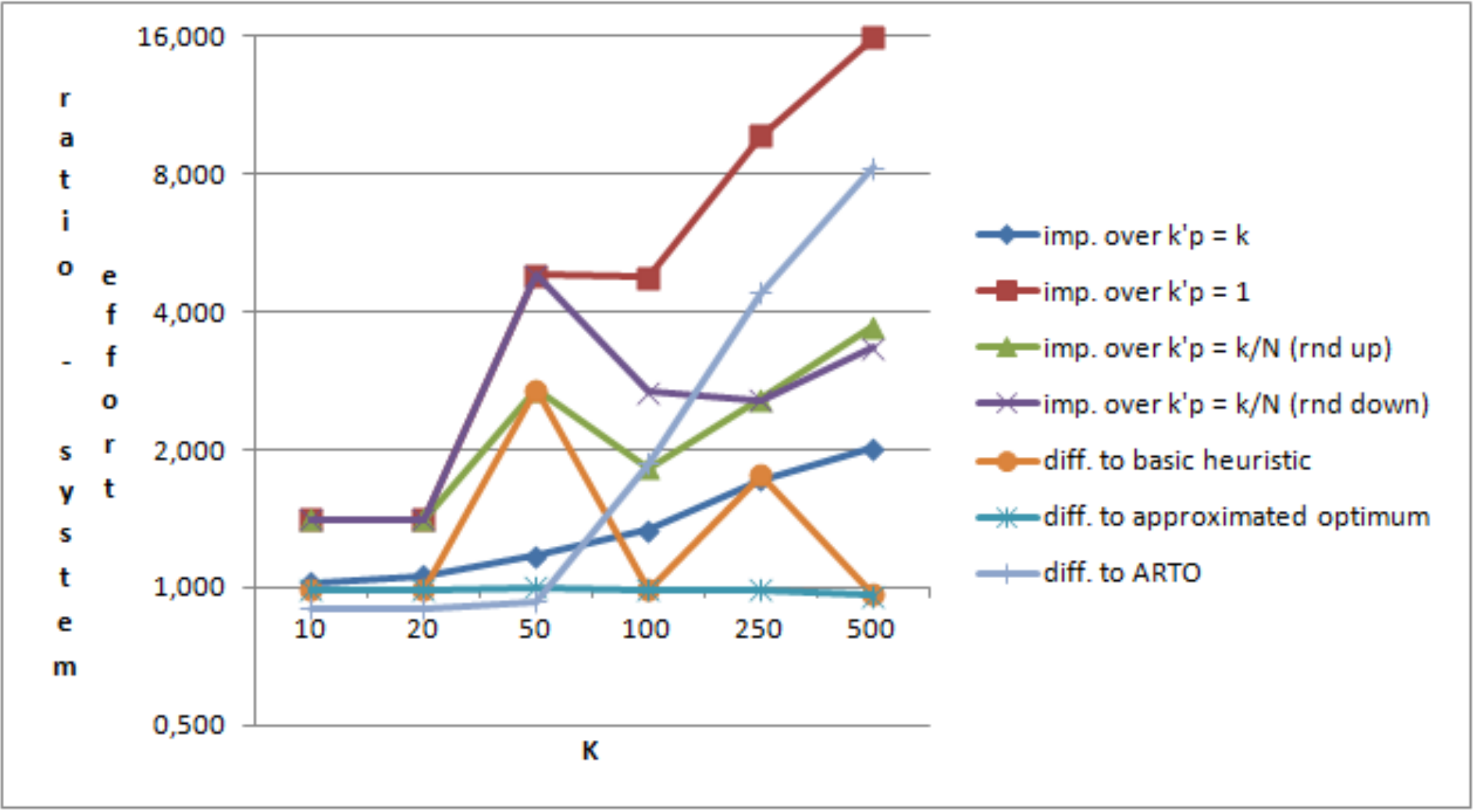}
	\caption{Ratio for system effort between \textit{enhanced heuristics}, approximated optimum, ARTO and five different approaches to determine the fetch size $k'_p$ in a peer to peer system with 49 peers and varying search amount K and 4 restrictions on census data.}%
	\label{chap:ADiT-fig:ratio_sysEff_49peers_04cons_cens}%
	\end{minipage}%
	\qquad
	\parbox{10cm}{
		\includegraphics[width=10cm]{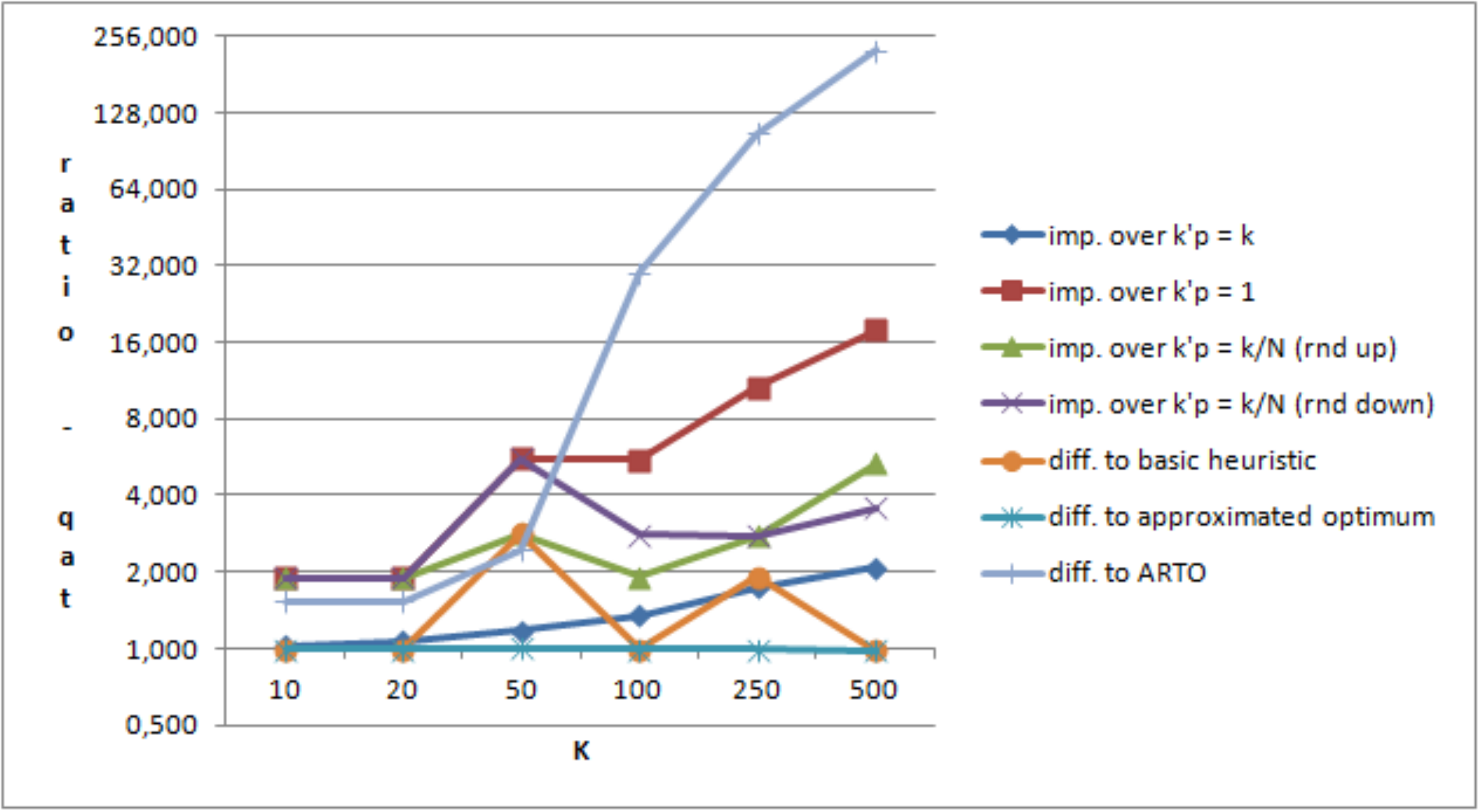}
	\caption{Ratio for query response time between \textit{enhanced heuristics}, approximated optimum, ARTO and five different approaches to determine the fetch size $k'_p$ in a peer to peer system with 49 peers and varying search amount K and 4 restrictions on census data.}%
	\label{chap:ADiT-fig:ratio_qat_49peers_04cons_cens}}%
\end{figure*}

In figure \ref{chap:ADiT-fig:ratio_sysEff_49peers_12cons_cens} and figure \ref{chap:ADiT-fig:ratio_qat_49peers_12cons_cens} we can see the $Ratio_{SE}$ and the $Ratio_{QAT}$ for a query with 12 restrictions in a peer to peer network with 49 peers storing census data. In these two figures we can observe the situation where the enhanced heuristics needs more iterations than the heuristics fetching $k'_p = k$ objects. This situation only occurred once in all of the test cases. Additionally, we see the same effect as in figure \ref{chap:ADiT-fig:ratio_sysEff_49peers_04cons_cens} and figure \ref{chap:ADiT-fig:ratio_qat_49peers_04cons_cens}, i.e. the curves are very similar but the  $Ratio_{QAT}$ is a little higher than $Ratio_{SE}$. When comparing figures \ref{chap:ADiT-fig:ratio_sysEff_49peers_12cons_cens} and \ref{chap:ADiT-fig:ratio_qat_49peers_12cons_cens} with figures \ref{chap:ADiT-fig:ratio_sysEff_49peers_04cons_cens} and \ref{chap:ADiT-fig:ratio_qat_49peers_04cons_cens} we can observe that the magnitude of the ratios is almost independent of the amount of restrictions. Furthermore, we observe in figure \ref{chap:ADiT-fig:ratio_sysEff_49peers_12cons_cens} and figure \ref{chap:ADiT-fig:ratio_qat_49peers_12cons_cens} that the ratios for ARTO increases at the point where the search amount exceeds the amount of peers in the network. This shows that it is better to ask each peer for more than only one object even when calling them sequentially.

\begin{figure*}
\centering
	\begin{minipage}{10cm}
		\includegraphics[width=10cm]{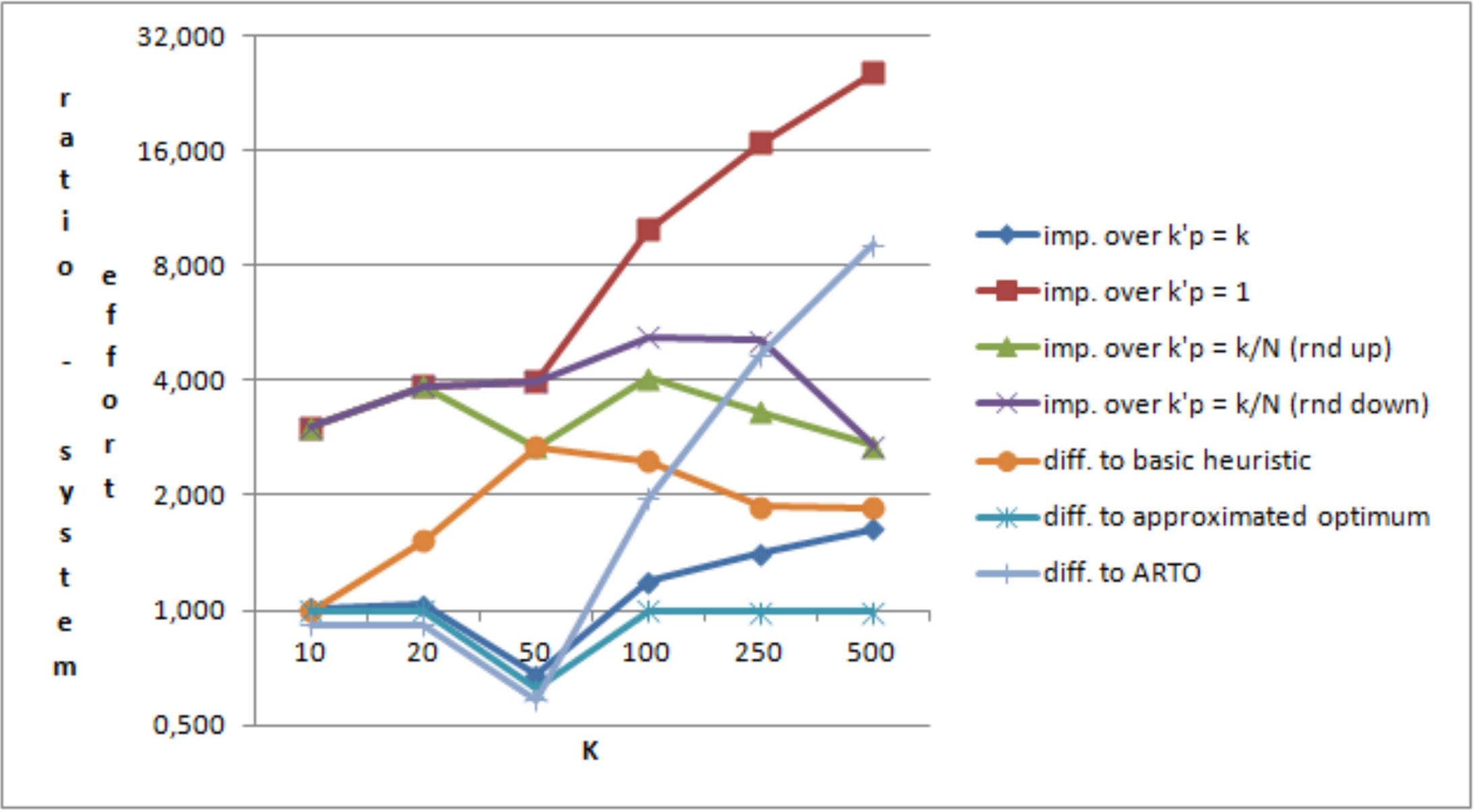}
	\caption{Ratio for system effort between \textit{enhanced heuristics}, approximated optimum, ARTO and five different approaches to determine the fetch size $k'_p$ in a peer to peer system with 49 peers and varying search amount K and 12 restrictions on census data.}%
	\label{chap:ADiT-fig:ratio_sysEff_49peers_12cons_cens}%
	\end{minipage}%
	\qquad
	\parbox{10cm}{
		\includegraphics[width=10cm]{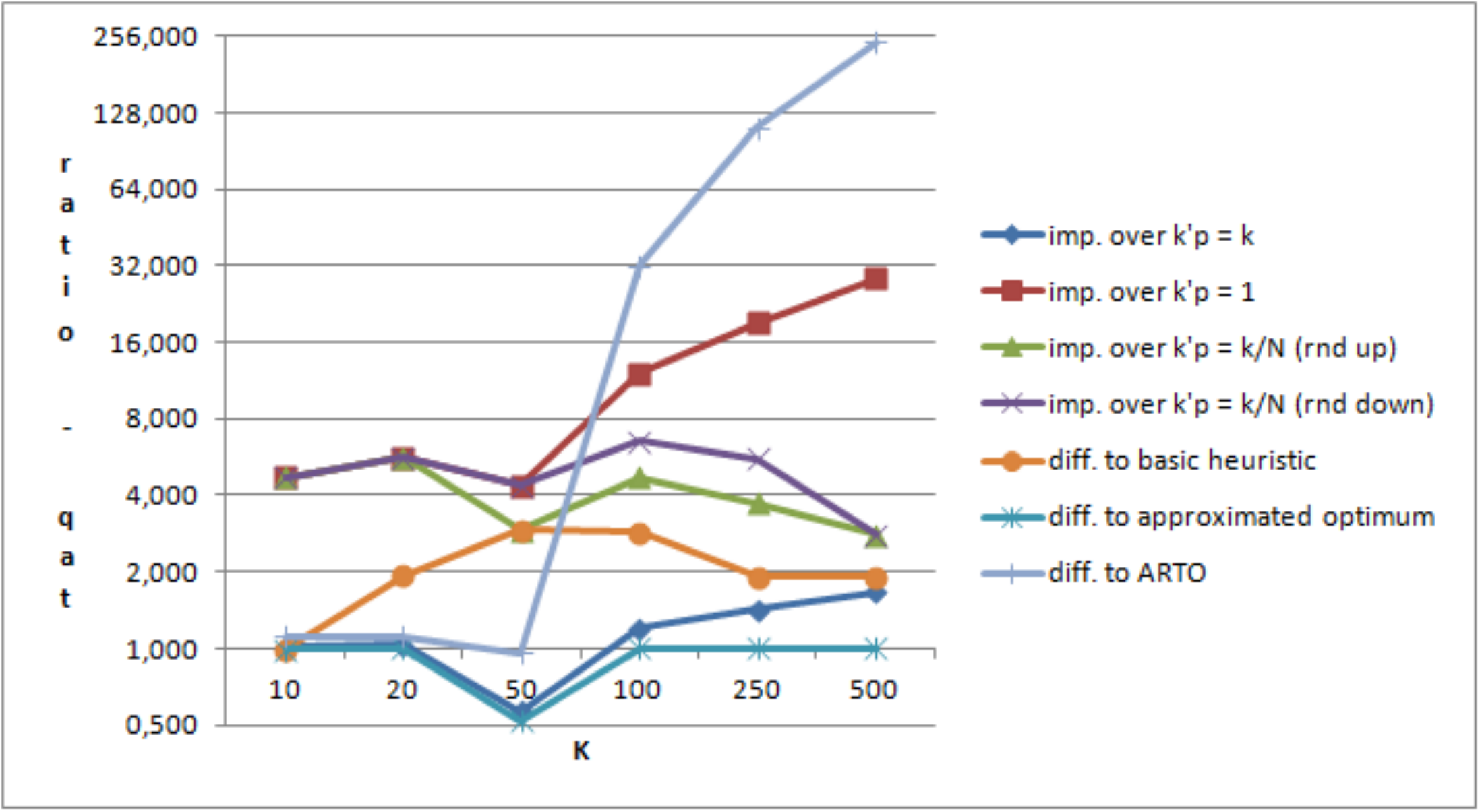}
	\caption{Ratio for query response time between \textit{enhanced heuristics}, approximated optimum, ARTO and five different approaches to determine the fetch size $k'_p$ in a peer to peer system with 49 peers and varying search amount K and 12 restrictions on census data.}%
	\label{chap:ADiT-fig:ratio_qat_49peers_12cons_cens}}%
\end{figure*}

The most important observations gathered through the performed test runs on random and US Census data are:
\begin{enumerate}
	\item ADiT is up to 200 times faster than ARTO in case the search amount gets higher than the amount of peers in the network.
	\item The system effort caused by ADiT is up to 8 times lower than the system effort caused by ARTO in case the search amount gets higher than the number of peers in the network.		
	\item The query response time of ARTO is in most cases worse than the query response time achieved with any of the presented ADiT heuristics.
\end{enumerate}

Additionally, we found some characteristics appearing in almost all test runs:
\begin{itemize}	
	\item The enhanced heuristics is close to the approximated optimum gathered through the extensive tests on the \textit{US Census Data (1990) Data Set}.
	
	\item The enhanced heuristics is better than all other presented heuristics, except in one single query (see figure \ref{chap:ADiT-fig:ratio_sysEff_49peers_12cons_cens} and figure \ref{chap:ADiT-fig:ratio_qat_49peers_12cons_cens}).
	
	\item The enhanced heuristics is between 2 and 32 times faster than the heuristics always fetching \textit{1 object} from each peer in parallel.
	
	\item The enhanced heuristics is about 3 to 8 times faster than heuristics fetching $\left \lceil \frac{k}{N} \right \rceil$ or $\left \lfloor \frac{k}{N} \right \rfloor$ objects from each peer in parallel.
	
	\item The enhanced heuristics is between 1.5 and 2.5 times faster than the heuristics fetching \textit{k objects} from each peer in parallel.
				
	\item The basic heuristics and the heuristics fetching \textit{k objects} from each peer in parallel turned out to be better than the other heuristics.
			
	
		
\end{itemize}

\section{Conclusion}
\label{chap:ADiT-sec:Conclusion}
We discussed distributed top-$k$ query processing from a new perspective. We motivated the need for an adaptive distributed top-$k$ query processing approach (short ADiT) and defined two goal measures, namely (1) the system effort and (2) the query response time.  Based on  data gathered through extensive experiments we derived a heuristics which can be used to determine a separate fetch size for each peer. We tested the developed heuristics with a large real data set, namely the US Census Data. In these tests we compared the enhanced heuristics against other heuristics and against ARTO \cite{Ryeng:2011:EDT:1997251.1997277}. We could show that ADiT can accelerate the query response time  and reduce the consumption of system resources significantly. Furthermore, we saw that the \textit{enhanced heuristics} is in most cases close to the best system effort and query response time approximately determined upfront. Additionally, we found that a heuristics fetching more objects is usually the better choice since searching and transmitting a few more objects has much lower costs than sending an additional request. Last but not least the gains achieved with ADiT increase with the size of the peer to peer network  and the number of requested results $k$.


\begin{thebibliography}{10}

\bibitem{UCICensusData}
{UCI Machine Learning Repository, US Census Data 1990
  \url{http://archive.ics.uci.edu/ml/datasets/US+Census+Data+(1990)}}, 2012.

\bibitem{Akbarinia:2006:RNT:1136637.1136639}
R.~Akbarinia, E.~Pacitti, and P.~Valduriez.
\newblock {Reducing network traffic in unstructured P2P systems using Top-$k$
  queries}.
\newblock {\em Distrib. Parallel Databases}, 19:67--86,  2006.

\bibitem{BPA2}
R.~Akbarinia, E.~Pacitti, and P.~Valduriez.
\newblock {Best position algorithms for top-$k$ queries}.
\newblock In {\em Proc. VLDB '07, 2007}.

\bibitem{Balke2005}
W.-T. Balke, W.~Nejdl, W.~Siberski, and U.~Thaden.
\newblock {Progressive Distributed Top-$k$ Retrieval in Peer-to-Peer Networks}.
\newblock In {\em Proc. ICDE '05, IEEE
  CS, 2005.}

\bibitem{Bruno:2002:TKS:568518.568519}
N.~Bruno, S.~Chaudhuri, and L.~Gravano.
\newblock {Top-$k$ selection queries over relational databases: Mapping
  strategies and performance evaluation}.
\newblock {\em ACM Trans. Database Syst.}, 27:153--187,  2002.

\bibitem{Conner2007}
W.~Conner, S.-w. Hwang, and K.~Nahrstedt.
\newblock {Unified Framework for Top-$k$ Query Processing in Peer-to-Peer
  Networks}.
\newblock 2007.

\bibitem{phddabringer}
C.~Dabringer.
\newblock {\em Efficient Local and Distributed Query Processing in a Biomedical
  Environment}.
\newblock PhD thesis, Alpen Adria Universit\"at Klagenfurt, 2012.

\bibitem{Dabringer:2011:ETR:1982185.1982414}
C.~Dabringer and J.~Eder.
\newblock {Efficient top-$k$ retrieval for user preference queries}.
\newblock In {\em Proc. ACM Symposium on Applied Computing},
  SAC '11, ACM, 2011.

\bibitem{Dabringer:2011:FTQ:2033546.2033563}
C.~Dabringer and J.~Eder.
\newblock {Fast top-$k$ query answering}.
\newblock In {\em Proc.  22nd int. conf. on Database
  and expert systems appl. -Part II}, DEXA'11, Springer-Verlag, 2011.

\bibitem{Eder:2009:ISF:1616930.1616937}
J.~Eder, C.~Dabringer, M.~Schicho, and K.~Stark.
\newblock {\em {Information Systems for Federated Biobanks}},
\newblock TLDKS Vol 1(1), Springer-Verlag, 2009.

\bibitem{fang2014efficient}
Q.~Fang and G.~Yang.
\newblock Efficient top-$k$ query processing algorithms in highly distributed
  environments.
\newblock {\em Journal of Computers}, 9(9):2000--2006, 2014.

\bibitem{Fang:2010:BPA:1917832.1918852}
Q.~Fang, Y.~Zhao, G.~Yang, B.~Wang, and W.~Zheng.
\newblock {Best Position Algorithms for Top-$k$ Query Processing in Highly
  Distributed Environments}.
\newblock In {\em Proc 1st Int. Conf.
  Networking and Distributed Computing}, ICNDC '10,  IEEE Computer Society, 2010.

\bibitem{Feuerstein:1999:OPP:555010}
S.~Feuerstein.
\newblock {\em {Oracle PL/SQL Programming Guide to Oracle8i Features}}.
\newblock O'Reilly \& Associates, 1999.

\bibitem{PLSQL}
S.~Feuerstein and B.~Pribyl.
\newblock {\em {Oracle PL/SQL programming}}.
\newblock  2009.

\bibitem{Frank+Asuncion:2010}
A.~Frank and A.~Asuncion.
\newblock {UCI Machine Learning Repository}, 2010.

\bibitem{Frank}
H.~Frank and J.~Eder.
\newblock {Towards an automatic integration of statecharts}.
\newblock In {\em Proc. ER'99, 1999}.

\bibitem{Hagihara:2009:MPM:1590953.1590977}
R.~Hagihara, M.~Shinohara, T.~Hara, and S.~Nishio.
\newblock {A Message Processing Method for Top-$k$ Query for Traffic Reduction in
  Ad Hoc Networks}.
\newblock In {\em Proc. Int. Conf. on
  Mobile Data Management}, MDM '09, IEEE Computer Society, 2009.

\bibitem{962155}
V.~Hristidis and Y.~Papakonstantinou.
\newblock {Algorithms and applications for answering ranked queries using
  ranked views}.
\newblock {\em The VLDB Journal}, 13(1):49--70, 2004.

\bibitem{1325952}
M.~Hua, J.~Pei, A.~W.~C. Fu, X.~Lin, and H.-F. Leung.
\newblock {Efficiently answering top-$k$ typicality queries on large databases}.
\newblock In {\em Proc. VLDB} 2007.

\bibitem{1391730}
I.~F. Ilyas, G.~Beskales, and M.~A. Soliman.
\newblock {A survey of top-$k$ query processing techniques in relational database
  systems}.
\newblock {\em ACM Comput. Surv.}, 40(4):1--58, 2008.

\bibitem{1272749}
N.~Mamoulis, M.~L. Yiu, K.~H. Cheng, and D.~W. Cheung.
\newblock {Efficient top-$k$ aggregation of ranked inputs}.
\newblock {\em ACM Trans. Database Syst.}, 32(3):19, 2007.

\bibitem{Owens:1998:BID:272975}
K.~T. Owens.
\newblock {\em {Building intelligent databases with Oracle PL/SQL, Triggers,
  and stored procedures (2nd ed.)}}.
\newblock Prentice-Hall,  1998.

\bibitem{MS_TopK}
C.~Re, N.~Dalvi, and D.~Suciu.
\newblock {Efficient Top-$k$ Query Evaluation on Probabilistic Data}.
\newblock {\em Data Engineering,
  2007.}


\bibitem{Ryeng:2011:EDT:1997251.1997277}
N.~H. Ryeng, A.~Vlachou, C.~Doulkeridis, and K.~N{\o}rv{\aa}g.
\newblock {Efficient distributed top-$k$ query processing with caching}.
\newblock In {\em Proc.  16th Int. Conf. on Database
  systems for advanced applications: DASFAA'11, Springer-Verlag,  2011.}

\bibitem{Vlachou:2008:ETQ:1376616.1376692}
A.~Vlachou, C.~Doulkeridis, K.~N{\o}rv{\aa}g, and M.~Vazirgiannis.
\newblock {On efficient top-$k$ query processing in highly distributed
  environments}.
\newblock In {\em Proc. 2008 ACM SIGMOD Int. Conf.
  on Management of Data}, ACM press, 2008.

\bibitem{OPT}
Z.~Zhang, S.-w. Hwang, K.~C.-C. Chang, M.~Wang, C.~A. Lang, and Y.-c. Chang.
\newblock {Boolean + ranking: querying a database by k-constrained
  optimization}.
\newblock In {\em Proc.  2006 ACM SIGMOD Int. Conf.
  on Management of Data}, ACM press, 2006.

\end{thebibliography}
\bibliographystyle{abbrv}

\end{document}